\documentclass[conference]{IEEEtran}
\usepackage{listings}

\ifCLASSINFOpdf
\else
\fi

\usepackage{graphicx} 
\begin{document}
%
\title{Bag-of-Features Image Indexing and Classification in Microsoft SQL Server Relational Database}

\author{\IEEEauthorblockN{Marcin Korytkowski, Rafa{\l} Scherer, Pawe{\l} Staszewski, Piotr Woldan}
\IEEEauthorblockA{Institute of Computational Intelligence\\ Cz\c{e}stochowa University of Technology\\
al. Armii Krajowej 36, 42-200 Cz\c{e}stochowa, Poland\\
Email: marcin.korytkowski@iisi.pcz.pl, rafal.scherer@iisi.pcz.pl}}


%


\IEEEoverridecommandlockouts
\IEEEpubid{\makebox[\columnwidth]{ 978-1-4799-8322-3/15/\$31.00~\copyright~2015~IEEE} 
\hspace{\columnsep}\makebox[\columnwidth]{ }}
\maketitle

\begin{abstract}
This paper presents a novel relational database architecture aimed to visual objects classification and retrieval. The framework is based on the bag-of-features image representation model combined with the Support Vector Machine classification and is integrated in a Microsoft SQL Server database.
\end{abstract}



%
\IEEEpeerreviewmaketitle

\section{Introduction}
\label{sec:intro}
Thanks to content-based image retrieval (CBIR) \cite{GuimaraesPedronette201491}\cite{DrozdaSG13}\cite{Kanimozhi20151099}\cite{Karakasis201522}\cite{Lin20146611}\cite{Liu2013188}\cite{Liu2012744}\cite{Rashedi201385}  we are able to search for similar images and classify them \cite{JAISCR-optimized}\cite{JAISCR-improved}\cite{JAISCR-breast}\cite{Shrivastava2014212}\cite{5206757_CVPR2009}. Images can be analyzed based on color representation \cite{Huang1997correlograms}\cite{Kiranyaz:2010}\cite{PassZabih1996}, textures \cite{ChangTexture1993}\cite{FrancosUnifiedTexture1993}\cite{JainGabor1991}\cite{smietanski2010texture}, shape \cite{Jagadish:1991}\cite{Kauppinen1995}\cite{Veltkamp:2000Shape} or edge detectors \cite{Zitnick:2014}. Recently, local invariant features have gained a wide popularity \cite{SIFT:Lowe:2004}\cite{Matas2004761}\cite{Mikolajczyk2004}\cite{Nister:2006}\cite{SivicVideoGoogle}. 
The most popular local keypoint detectors and descriptors are SURF \cite{SURF:Bay:2008}, SIFT \cite{SIFT:Lowe:2004} or ORB \cite{Rublee2011}. 
To find similar images to a query image, we need to compare all feature descriptors of all images usually by some distance measures. Such comparison is enormously time consuming and there is ongoing worldwide research to speed up the process. Yet, the current state of the art in the case of high-dimensional computer vision applications is not fully satisfactory. The literature presents  countless methods and variants utilizing e.g. a voting scheme or histograms of clustered keypoints. 
They are mostly based on some form of approximate search. 
Recently, the bag-of-features (BoF)  approach \cite{Grauman2005}\cite{Philbin2007}\cite{SivicVideoGoogle}\cite{SlavaBoF}\cite{ZhangMarszalek2006} has  gained in popularity. In the BoF method, clustered vectors of image features are collected and sorted by the count of occurrence (histograms). All individual descriptors or approximations of sets of descriptors presented in the histogram form must be compared. Such calculations are computationally expensive. Moreover, the BoF approach requires to redesign the classifiers when new visual classes are added to the system. 

The paper deals with a visual query-by-example problem in relational databases. Namely, we developed a system based on  Microsoft SQL Server which is able to classify a sample image or to return similar images to this image. 
Storing huge amount of undefined and unstructured binary data and its fast and efficient searching and retrieval is the main challenge for database designers. Examples of such data are images, video files etc. Users of world most popular relational database management systems (RDBMS) such as Oracle, MS SQL Server and IBM DB2 Server are not encouraged to store such data directly in the database files. The example of such an approach can be Microsoft SQL Server where binary data is stored outside the RDBMS and only the information about the data location is stored in the database tables. MS SQL Server utilizes a special field type called FileStream which integrates SQL Server database engine with NTFS file system by storing binary large object (BLOB) data as files in the file system. Microsoft SQL dialect (Transact-SQL) statements can insert, update, query, search, and back up FileStream data. Application Programming Interface provides streaming access to the data. FileStream uses operating system cache for caching file data. This helps to reduce any negative effects that FileStream data might have on the RDBMS performance. 
Filestream data type is stored as a \texttt{varbinary(max)}  column with pointer to actual data which are stored as BLOBs in the NTFS file system. By setting the FileStream attribute on a column and consequently storing BLOB data in the file system, we achieve the following advantages:
\begin{itemize}
\item performance is the same as the NTFS file system and SQL Server cache is not burden with the Filestream data,
\item Standard SQL statements such as SELECT, INSERT, UPDATE, and DELETE work with FileStream data; however, associated files can be treated as standard NTFS files.
\end{itemize}
In the proposed system, large image files are stored in a FileStream field. Unfortunately, despite using this technique, there does not exist a technology for fast and efficient retrieval of images based on their content in existing relational database management systems. 
 Standard SQL language does not contain commands for handling multimedia, large text objects, and spatial data.

We designed a special type of field, in which a set of keypoints can be stored in an optimal way, as so-called User-Defined Type (UDT). Along with defining the new type of field, it is necessary to implement methods to compare its content. When designing UDT, various features must be also implemented, depending on implementing the UDT as a class or a structure, as well as on the format and serialization options. This could be done using one of the supported .NET Framework programming languages and the UDT can be implemented as a dynamic-link library (DLL), loaded in MS SQL Server. Another major challenge is to create a special database indexing algorithm, which would significantly speed up answering to SQL queries for data based on the newly defined field.
As aforementioned, standard SQL does not contain commands for handling multimedia, large text objects and spatial data. Thus, communities that create software for processing such specific data types, began to draw up SQL extensions, but they transpired to be incompatible with each other. That problem caused abandoning new task-specific extensions of SQL and a new concept won, based on libraries of object types SQL99 intended for processing specific data applications. The new standard, known as SQL/MM (full name: SQL Multimedia and Application Packages), was based on objects, thus programming library functionality is naturally available in SQL queries by calling library methods.  SQL/MM consists of several parts: framework -- library for general purposes, full text -- defines data types for storing and searching large amount of text, spatial -- for processing geospatial data, still image -- defines types for processing images and data mining -- data exploration. 
There are also attempts to create some SQL extensions using fuzzy logic for building flexible queries.  In \cite{Dubois2001} possibilities of creating flexible queries and queries based on user’s examples are presented.  It should be emphasized that the literature shows little efforts of creating a general way of querying multimedia data.

The main contribution and novelty of the paper is as follows:
\begin{itemize}
\item  We present a novel system for content-based image classification built in a Microsoft SQL Server database,
\item  We created a special database indexing algorithm, which will significantly speed up answering to visual query-by-example SQL queries in relational databases.
\end{itemize}

The paper is organized as follows. Section \ref{sec:system} describes the proposed database system. Section \ref{sec:simul} provides simulation results on the the PASCAL Visual Object Classes (VOC) 2012 dataset \cite{Everingham10}. 
%
%
%
%
%
%
%
%
%
%
%
\section{System Architecture and Relational Database Structure}
\label{sec:system}
Our system and generally BoF can work with various image features. In the paper we use SIFT features as an example. To calculate SIFT keypoints we used the OpenCV library. We did not use functions from this library as a user defined functions (UDF) directly in the database environment because:
\begin{enumerate}
\item User Defined Functions can be written only in the same .NET framework version as the MS SQL Server (e.g. MS SQL Server was created based on .NET 4.0)
\item Calculations used to find image keypoints are very complex, thus running such computations directly on the database server causes the database engine to become unresponsive. 
\end{enumerate}
\begin{figure}
\label{fig:system}
\includegraphics[width=0.48\textwidth]{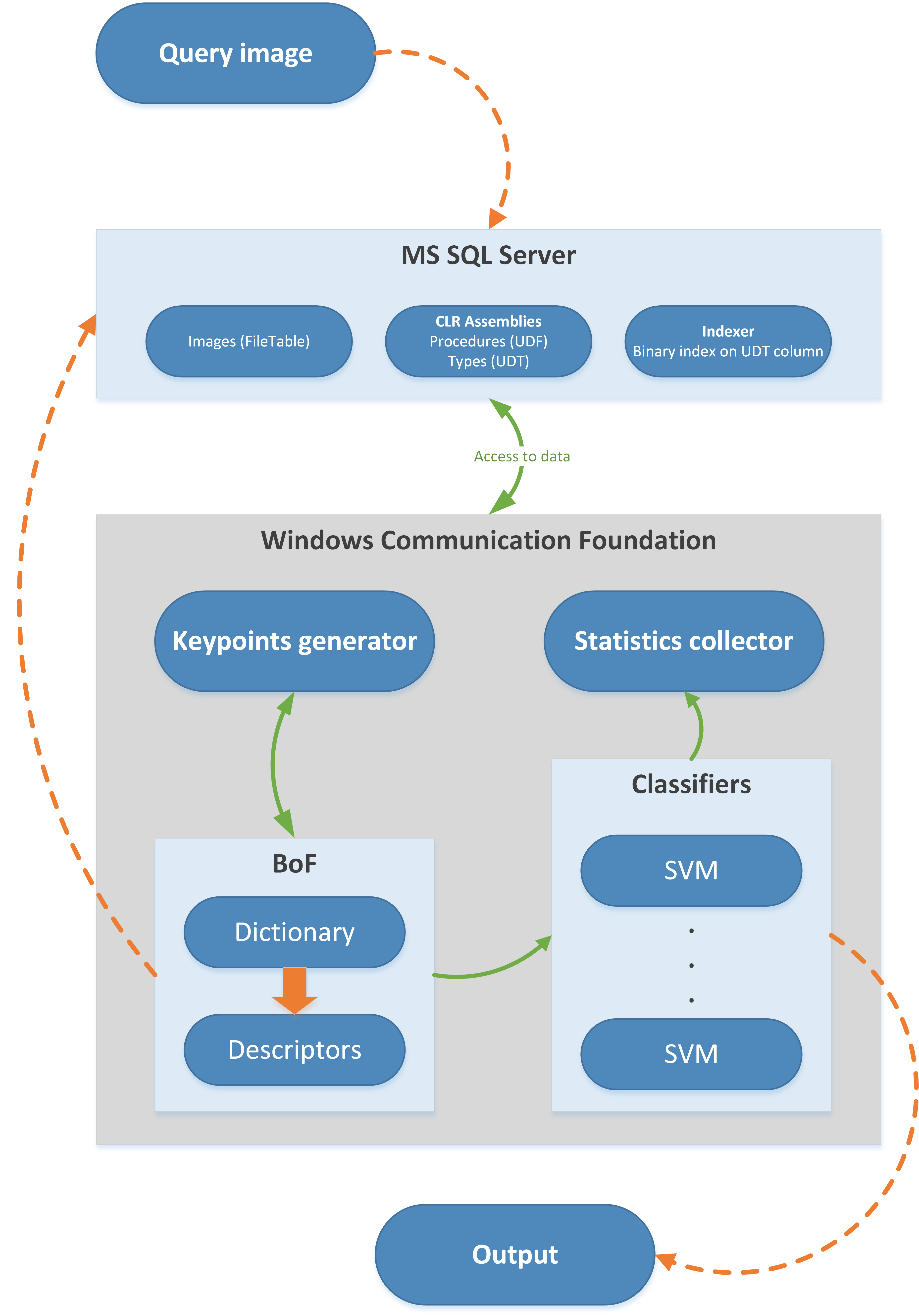}
\caption{System architecture}
\end{figure}

From the above-mentioned reasons, similarly as in the case of the Full Text Search technology, the most time-consuming computations are moved to the operating system as background system services of WCF (Windows Communication Foundation). 
WCF Data Service works as the REST architecture (Representational State Transfer) which was introduced by Roy T. Fielding in his PhD thesis \cite{fielding2000architectural}. Thanks to WCF technology, it is relatively easy to set the proposed solution in the  Internet. 
To store image local keypoints in the database, we created a User Defined Type  (column \texttt{sift\_keypoints} in \texttt{SIFTS} table). These values are not used in the classification of new query images. They are stored in case we need to identify a new class of objects in the existing images as having keypoint values, we would not have to generate keypoint descriptors again. Newly created type was created in C\# language as a CLR class and  only its serialized form is stored in the database. 
\begin{figure*}
\label{fig:fig_datab}
\includegraphics[width=0.9\textwidth]{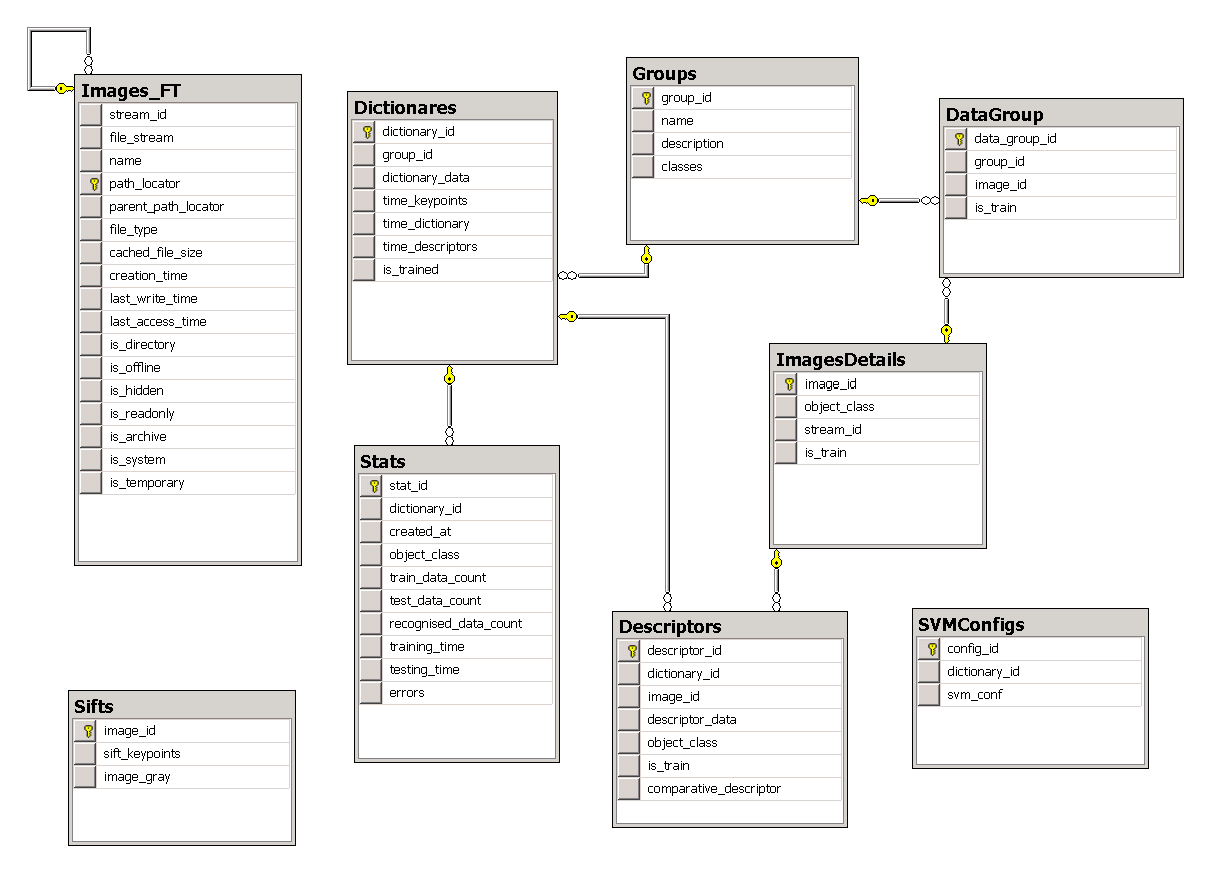}
\caption{Proposed database structure}
\end{figure*}
The database stores also Support Vector Machine classifiers parameters in the \texttt{SVMConfigs}  table. Such an approach allows running any time the service with learned parameters. Running the service in the operating system will cause reading SVM classifiers from the database. The \texttt{Stats} table is for collecting algorithm statistics, where the most important numbers are execution times for consecutive stages of the algorithm. The \texttt{Images} table is for storing membership of images for visual classes. \texttt{Dictionaries} table is responsible for storing keypoint clusters data, and these cluster parameters are stored in the \texttt{DictionaryData} field of UDT type:
\lstset{language=C++}  
\begin{lstlisting}

public struct DictionaryData: 
         INullable, IBinarySerialize
{
    private bool _null;
    public int WordsCount {get; set;}
    public int SingleWordSize {get; set;}
    public double[][] Values {get; set;}
    public override string ToString()
    ...
}
\end{lstlisting}
The \texttt{WordsCount} variable stores information about the number of words in the BoF dictionary, \texttt{SingleWordSize} variable value depends on the algorithm used to generate image keypoint descriptors, and in case of the SIFT algorithm, it equals 128. Two-dimensional matrix \texttt{Values} stores information regarding cluster centers. 
The system operates in two modes:
\subsubsection{learning mode}
Image keypoint descriptors are clustered to build a bag-of-features dictionary by the $k$-means algorithm. Cluster parameters are stored in \texttt{DictionaryData} variables. Next, image descriptors are created for subsequent images. They can be regarded as histograms of membership of image local keypoints to words from dictionaries. We use \texttt{SIFTDetector} method from the Emgu CV (http://www.emgu.com) library with the following signature: \texttt{ComputeDescriptorsRaw(Image<Gray, byte>grayScaleImage, Image<Gray, byte> mask , VectorOfKeypoint keypoints)}. 
Obtained descriptors are then stored in the \texttt{Descriptors} table
of UDT type:
\lstset{language=C++}  
\begin{lstlisting}
public struct DescriptorData: 
             INullable, IBinarySerialize
{
    //  Private member
    private bool _null;
    public int WordsCount {get; set;}
    public double[] Values {get; set;}
    ...
}
\end{lstlisting}
Using records from this table, learning datasets are generated for SVM classifiers to recognize various visual classes. These classifiers parameters are stored after the training phase in the \texttt{SVMConfigs} table. 
\subsubsection{Classification Mode}
In the classification phase, the proposed system works fully automatically. After sending an image file to the \texttt{Images\_FT} table, a service generating local interest points is launched. In the proposed approach, we use SIFT descriptors. Next, the visual descriptors are checked against membership to clusters stored in the database in the \texttt{Dictionaries} table and on this base, the histogram descriptor is created. To determine membership to a visual class we have to use this vector as the input for all SVM classifiers obtained in the learning phase. For the classification purposes, we extended SQL language and defined \texttt{GetClassOfImage()} method in C\# language and added it to the set of User Defined Functions. The argument of this method is the file identifier from the FileTable table. 

Microsoft SQL Server constraints the sum of indexed columns to 900 bytes. Therefore, it was not possible to create an index on the columns constituting visual descriptors. To allow fast image searching of the \texttt{Descriptors} table, we created a field \texttt{comparative\_descriptor } that stores descriptor value  hashed by the MD5 algorithm. It allowed creating index on this new column, thus the time to find an image corresponding with the query image was reduced substantially. 
\section{Numerical Simulations}
\label{sec:simul}
We tested the proposed method on three classes of visual objects taken from the PASCAL Visual Object Classes (VOC) dataset \cite{Everingham10}, namely: Bus, Cat and Train. We divided these three classes of objects into learning and testing examples. The testing set consists of 15\% images from the whole dataset. Before the learning procedure we generated local keypoint vectors for all images from the Pascal VOC dataset using the SIFT algorithm. 
All simulations were performed on a Hyper-V virtual machine with MS Windows Operating System (8 GB RAM, Intel Xeon X5650, 2.67 GHz). The testing set only contained  images that had never been presented to the system during learning process.

The bag-of-features image representation model combined with the Support Vector Machine (SVM) classification was run five times for various dictionary sizes: 40, 50, 80, 100, 130 and 150 words. Dictionaries for the BoF were created using C++ language, based on the OpenCV Library \cite{bradski2000opencv}.  
The results of the BoF and SVM classification on the testing data are presented in Table I. The SQL queries responses are nearly real-time for even relatively large image datasets. 
\begin{figure}
\label{fig:zdj}
\includegraphics[width=0.48\textwidth]{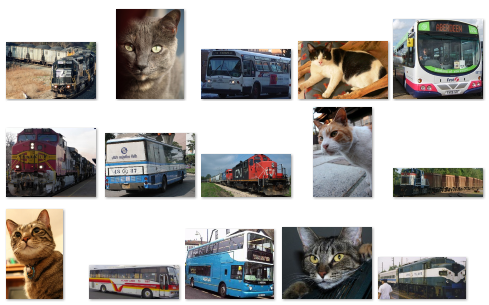}
\caption{Exemplary images from testing dataset}
\end{figure}
\begin{table}
\label{tab:res}
\caption{Numerical simulation results for various BoF dictionary size.}
\begin{tabular}{|p{0.4in}|p{0.3in}|p{0.3in}|p{0.3in}|p{0.3in}|p{0.3in}|p{0.3in}|} \hline 
\textbf{Words:} & \textbf{40} & \textbf{50} & \textbf{80} & \textbf{100} & \textbf{130} & \textbf{150} \\ \hline 
Bus & 40\% & 50\% & 60\% & 60\% & 70\% & 50\% \\ \hline 
Cat & 90\% & 80\% & 50\% & 80\% & 80\% & 80\% \\ \hline 
Train & 0\% & 0\% & 10\% & 20\% & 10\% & 10\% \\ \hline 
\textbf{Result:} & \textbf{43\%} & \textbf{43\%} & \textbf{40\%} & \textbf{53\%} & \textbf{53\%} & \textbf{47\%} \\ \hline 
\end{tabular}
\end{table}
\section{Conclusion}
We presented a method that allows integrating relatively fast content-based image classification algorithm with relational database management system. Namely, we used bag of features, Support Vector Machine classifiers and special Microsoft SQL Server features, such as User Defined Types and CLR methods, to classify and retrieve visual data. Moreover, we created indexes to search for the same query image in large sets of visual records. Described framework allows automatic searching and retrieving images on the base of their content using the SQL language. The SQL responses are nearly real-time on even relatively large image datasets. 
The system can be extended to use different visual features or to have a more flexible SQL querying command set. 
%
%
%

\section*{Acknowledgment}
This work was supported by the Polish National Science Centre (NCN)  under project number DEC-2011/01/D/ST6/06957.



%
%
%

\bibliography{CBIR_RS}
\bibliographystyle{IEEEtran}

\end{document}